\documentclass[a4paper,11pt]{article}
\pdfoutput=1 

\usepackage{jcappub} 

\usepackage[T1]{fontenc} 

\usepackage{lmodern}

\usepackage[utf8]{inputenc}
\usepackage{amsmath}
\usepackage{amssymb}
\usepackage{esint}
\usepackage{braket}

\makeatletter

\usepackage{graphics}

\usepackage{setspace}
\usepackage{graphicx}
\usepackage{color}
\usepackage{epstopdf}
\usepackage{setspace}
\usepackage{makeidx}

\usepackage{color}
\usepackage{caption}
\usepackage{subcaption}
\usepackage[export]{adjustbox}

\def\be{\begin{equation}}
\def\ee{\end{equation}}

\newcommand{\dint}{\int\hspace{-0.3cm}\int}
\newcommand{\nn}{\nonumber \\}

\title{\boldmath Quantum corrections for the cubic Galileon in the covariant language}

\author[a]{Ippocratis D. Saltas} \author[b]{and Vincenzo Vitagliano}

\affiliation[a]{Institute of Astrophysics and Space Sciences, Faculty of Sciences,\\ 
Campo Grande, PT1749-016 Lisboa, Portugal}
\affiliation[b]{Multidisciplinary Center for Astrophysics \& Department of Physics,\\ 
Instituto Superior T\'ecnico, University of Lisbon, \\Av. Rovisco Pais 1, 1049-001 Lisboa, Portugal}
\emailAdd{isaltas@fc.ul.pt}
\emailAdd{vincenzo.vitagliano@ist.utl.pt}
\abstract{We present for the first time an explicit exposition of quantum corrections within the cubic Galileon theory including the effect of quantum gravity, in a background-- and gauge--invariant manner, employing the field--reparametrisation approach of the covariant effective action at 1--loop. We show that the consideration of gravitational effects in combination with the non--linear derivative structure of the theory reveals new interactions at the perturbative level, which manifest themselves as higher--operators in the associated effective action, which' relevance is controlled by appropriate ratios of the cosmological vacuum and the Galileon mass scale. The significance and concept of the covariant approach in this context is discussed, while all calculations are explicitly presented. }

\begin{document}
\maketitle
\flushbottom
\section{Introduction} 
The yet unresolved problem of dark energy, as well as the need to provide a consistent framework for the primordial inflationary paradigm led to the exploration of scalar--field theories with non--trivial interactions and dynamics. Among the infinitely many options, of particular significance and attention have been theories leading to second--order equations, due to the well-known problem of Ostrogradski instability associated with higher--order dynamics \cite{wood}. The latter have been dubbed ``Galileon'' theories \cite{Horndeski:1974wa,Nicolis:2008in,Deffayet:2009wt}, and present with a family of effective scalar--field theories exhibiting non--trivial derivative interactions and coupling to gravity which have been re-discovered in distinct setups, such as in brane-world contexts and theories of massive gravity \cite{Dvali:2000hr,deRham:2010kj,deRham:2012az,deRham:2010eu}. 
The name ``Galileon'' stems from the fact that the theory enjoys the so--called Galileon symmetry,
\be
\phi(x) \to \phi(x) + b_\mu x^\mu + c,
\ee
which plays a central role in the various attractive classical and quantum properties of these theories. The presence of a potential term or some non--minimal coupling to gravity would break this symmetry, however in scenarios where the Galileon plays the role of the inflaton a potential term is usually required.
 
On the phenomenological side, the idea that the Galileon field could be responsible for accelerating the Universe at early-- or late--times has led to the exploration of a variety of cosmological phenomenology \cite{Chow:2009fm,Silva:2009km,Kobayashi:2010cm,DeFelice:2010nf,DeFelice:2010pv,Creminelli:2010ba,Gannouji:2010au,Kobayashi:2010wa,Kobayashi:2009wr,Burrage:2010cu,Germani:2016gzh}. At smaller scales, any gravitational theory introducing new degrees of freedom should agree with local tests through some mechanism suppressing potential fifth--force effects. Galileons achieve this through the so--called Vainshtein mechanism \cite{Vainshtein:1972sx}, which suppresses the scalar's effects sufficiently close to matter sources due to the dominance of non--linear derivative interactions, effectively switching off the scalar's fifth force and recovering standard gravity. 

The quantum--mechanical properties of Galileon theories, which will be the topic of the present work, have also attracted significant attention. The reason is twofold. On the one hand, the non--trivial interactions in the theory allow for a rich quantum--mechanical phenomenology and a theoretical framework for exploring potential novel features not exhibited by standard scalar--field matter, while on the other hand, issues such as the strong coupling scale and Vainshtein screening, as well as possible cosmological effects make the understanding of quantum corrections essential \cite{Kaloper:2014vqa,Pirtskhalava:2015nla}. One of the most important results in this regard is that that loops of the Galileon do not lead to renormalisation of  the Galileon interactions themselves, shown in \cite{Luty} for the case of the cubic Galileon and in \cite{Hinter, deRham:2014zqa,Heisenberg:2014raa,Brouzakis:2013lla,dePaulaNetto:2012hm, Amado} in more general setups. The striking property of Galileon theories in this regard is the fact that new operators generated by Galileon loops exhibit higher number of derivative interactions than those in the original bare action, leaving the original operators unrenormalised. All these results have been proven at the semi--classical level and in an effective--field theory sense, with the effect of graviton loops in this context being so far unknown. 

In this work, we will present for the first time an explicit exposition of the quantum corrections including the graviton loops, following our previous work presented in \cite{Saltas:2016nkg}. Most importantly, aiming for the least ambiguous result, our analysis will make use of a background and gauge invariant framework, in particular the method of the Vilkovisky--DeWitt effective action. It is well known that the standard calculation of the (off--shell) effective action is plagued by a dependence on the choice of background and gauge, potentially leading to ambiguous results. Although the issue of background dependence concerns both gauge and non gauge field theories, the presence of gravity introduces in principle an extra dependence on the choice of gauge. The method of DeWitt \cite{DeWitt:1967yk,DeWitt:1967ub,DeWitt:1967uc} and Vilkovisky \cite{Vilkovisky:1984st,gospel} provides a geometrical resolution to this problem that allows covariance in field-space to be preserved, this way ensuring gauge- and background-independence (see also \cite{Parker:2009uva} for an excellent review.) 
The original Vilkovisky--DeWitt method has been since then further generalised through a formidable work by DeWitt himself \cite{deW} to allow for the calculation of loops higher than one. However, it has been showed \cite{Rebhan:1986wp,Rebhan:1987cd} that  the refined approach and the original one coincide up to one loop. 
The covariant effective action of Vilkovisky and DeWitt has been employed before in a variety of settings ranging from the calculation of the running of gravitational couplings \cite{Toms:2008dq,Donkin:2012ud}, and scalar-tensor theories \cite{Mackay:2009cf, Pietrykowski:2012nc} as well as gravity coupled to electromagnetism \cite{Toms:2010vy, Toms:2009vd,Buchbinder:1992rb} and stability of electroweak potential and inflation \cite{Moss:2015gua}, while some particular technical aspects of it have been also discussed in \cite{Odintsov:1991yx}. 

For our analysis, we shall be focusing on a subset of the Galileon family, namely the cubic Galileon, which exhibits the essential non--trivial features of Galileon theories. The final result of our calculation will be the 1--loop effective action for the theory, in the presence of both scalar and graviton loops. As we will explain explicitly later on, the requirement of field--reparametrisation invariance, combined with the presence of the gravitational back-reaction, induces new non--trivial interactions in the scalar sector of the theory leading to a new operator structure, unnoticed in previous analyses. In particular, a key point in this regard is the fact that although the original bare action of the theory respects shift--symmetry, the Galileon field acquires a gravitationally--induced mass--type interaction of the order of the cosmological vacuum, giving rise to genuinely new, non--Planck suppressed operators at 1--loop. 

{\it Our analysis highlights two important features in this context. The first is the significance of quantum--gravitational corrections for theories with non--trivial derivative structure even at energies well--below the Planck scale, and on the same time suggests the importance of the background/gauge--invariant approach.} 

We structure the paper as follows: In Section \ref{sec:Galileon} we introduce the theory of the cubic Galileon, while in Section \ref{sec:Formalism} we lay down the fundamental principles underlying the formalism of the covariant effective action. Our main calculation and results are presented in Section \ref{sec:Calculation}, and we conclude in Section \ref{sec:Conclusions}. Some intermediate calculations are kept for the Appendix.

\section{Cubic Galileon theory} \label{sec:Galileon}

In this section we shall introduce the main action describing the theory under consideration, and the associated classical dynamics. The setup we will be considering is that of the cubic Galileon scalar--field theory minimally coupled to gravity described by the (Euclidean) action
\begin{align}
S= \int  d^{4}x \sqrt{g} \left( L_{\text{G}} + L_{\text{M}} \right) \equiv \int  d^{4}x \sqrt{g} \left[ -\frac{2}{\kappa^2} R +X\left(1 + B \right)+\frac{4\Lambda}{\kappa^2}\right], \label{action:0}
\end{align}
where $\kappa^2 \equiv 32 \pi G$, the scalar field's kinetic term defined as
\be
X \equiv (1/2) g^{\mu \nu}\nabla_\mu \phi \nabla_\nu \phi,
\ee
and 
\be
B \equiv  \Box \phi /M^3   \equiv g^{\mu \nu} \nabla_\mu \nabla_\nu \phi /M^3, 
\ee 
with $M$ here being an in principle arbitrary energy scale associated with the cubic Galileon term. We shall discuss its relevance later on, after we derive the 1--loop quantum corrections for the theory. $\Lambda$ denotes the cosmological constant, which will play an important role in the quantum analysis. Let us remind that, in pure scalar field theory the vacuum energy, represented by $\Lambda$ is physically irrelevant, however this is no longer true in the presence of gravity. 

The action (\ref{action:0}) is invariant under the Galileon symmetry up to total derivatives,
\be
\phi(x) \to \phi(x) + b_{\mu} x^{\mu} + c \label{Galileon-symmetry},
\ee
with $ b_{\mu}$ and $c$ arbitrary constants. Notice that, the presence of a potential term $V(\phi)$ for the Galileon field or a non--minimal coupling to curvature $\sim \xi \phi^2 R$ would break it, while for $b_\mu =0$, (\ref{Galileon-symmetry}) simply corresponds to a shift--symmetry transformation. 
The non--linear derivative interaction term $\sim X \Box \phi$ in (\ref{action:0}) corresponds to the third--order term (the first two being the linear $\sim \phi$ and quadratic $X$ terms respectively) from a total of five terms which respect the symmetry (\ref{Galileon-symmetry}), and are built out of the scalar and its derivatives. All of these terms give rise to second--order equations motion for the scalar. 
In particular, the theory described by the action (\ref{action:0}) is a particular case of the so--called kinetic--gravity braiding models, where on a curved (cosmological) background the dynamics of the scalar mix with those of the metric non--trivially, due to the coupling of metric's derivatives through the d'Alembertian with those of the scalar field \cite{Pujolas:2011he}. This mixing cannot be untwisted with a field redefinition, and poses a genuinely non--trivial feature of the theory.

The classical dynamics of the theory are found by varying the action (\ref{action:0}) with respect to $\phi$\footnote{Notice that, since we will be working on a flat background later on, we evaluate the classical equations on a flat metric for later convenience.},
\begin{align}
&\mathcal{E}_{\phi} \equiv  \frac{\delta S}{\delta \phi} = -M^{3} B(1 + B) + M^{-3}\partial_{\alpha} \partial_{\kappa} \phi   \cdot \partial^{\alpha} \partial^{\kappa}\phi, \label{EOM}
\end{align}
and the corresponding equation of motion follows as $\mathcal{E}_{\phi} = 0$. 
As expected, the scalar-field equations of motion are of second-order nature, despite the appearance of higher--order derivative terms in the action. 
In a similar fashion, the energy--momentum tensor for the scalar is calculated as
\begin{align}
& T_{\alpha \beta} \equiv   -\left. \frac{1}{\sqrt{g}} \frac{\delta S}{\delta g^{\alpha \beta}} \right|_{g_{\alpha \beta} = \eta_{\alpha \beta}} =  \frac{1}{2}L_{\text{M}} \eta_{\alpha \beta}  - \frac{1}{2}\partial_\alpha \phi \partial_\beta \phi \left( 1 + B \right) + M^{-3} \partial_{(\alpha} X \partial_{\beta)}\phi,  \label{EMT}
\end{align}
and its expression along with the scalar's equation of motion will be useful for the analysis to follow. In the case that the Galileon is responsible for the Universe's acceleration (with $\Lambda = 0$) the source scalar's energy--momentum tensor provides an effective cosmological fluid based on the dominance of derivative interactions. 

\section{Geometrical considerations and the covariant effective action} \label{sec:Formalism}
In this section we will introduce the necessary geometrical concepts and tools required to proceed with the evaluation of the 1--loop covariant effective action. We will start by discussing the significance of covariance in field space and proceed with the particular application to the cubic Galileon theory.

Before we start with the technical part of the discussion it is helpful to briefly comment on the issue of gauge and background dependence in the calculation of the effective action. In the standard evaluation of the path integral and the associated effective action, one of the most commonly employed approach is that of the background-field method, which accounts to splitting the physical expectation value of the field into a fixed-background and a fluctuating piece respectively. For gauge theories this presents with a breaking of local symmetries, since the fixed-background field does not transform simultaneously with the fluctuating piece, leading to results which are in principle dependent on the gauge choice. What is more, the use of the background field to define covariant derivatives and consequently momenta, leads to a further dependence on the choice of background. It becomes therefore crucial to ensure that any conclusions about the quantum dynamics of the theory are not plagued by some artificial effect associated with the choice of gauge and/or background (see for example Refs \cite{Falkenberg:1996bq,Labus:2016lkh,Ohta:2016npm,Ohta:2016jvw,Morris:2016nda}). The method of Vilkovisky-DeWitt which we shall briefly outline below and also employ in our calculation, suggests a way around to the above issues, through the introduction of an effective action which is covariant in the space of fields, allowing for field--reparametrisation invariance of the results.

To introduce the main idea of the formalism \footnote{Here, we will be closely following \cite{Parker:2009uva}.}, let us start with an abstract setup considering an action $S = S[\Phi^i]$, that depends on a set of fields $\Phi^i$, where $i$ is a generalised index labelling fields with arbitrary tensor structure. Notice also that, Greek letters will be denoting usual spacetime indices. Let us also assume that the action enjoys the local symmetry
\be
S[\Phi^i] = S[\Phi^{i}_\epsilon],
\ee
with the infinitesimal field transformation defined through
\be
\delta \Phi^i \equiv \Phi^{i}_\epsilon - \Phi^i = K_{\alpha}^{i}[\Phi^j] \delta \epsilon^\alpha.
\ee
We understand $\delta \epsilon^\alpha$ as the transformation vectors (``gauge vectors'') and $K_{\alpha}^{i}[\Phi^j]$ as the coordinates of the transformation (``the generators'') respectively. The theory we will be elaborating on, is constructed out of the metric and a scalar field, i.e $\Phi^i = \{g_{\mu \nu}, \phi \}$. The gauge symmetry of a gravitational theory is dictated by general covariance, i.e the local redefinition of coordinates,
\be
 x^\mu\rightarrow \widetilde{x}^\mu=x^\mu+\delta\epsilon^\mu(x),
\ee
under which the metric and scalar field transform according to
\begin{eqnarray}
 \delta g_{\mu\nu}^{\textrm{coor}}(x)&=&\int d^n x' K^{g_{\mu\nu}(x)}{}_\lambda(x,x')\delta\epsilon^\lambda(x'), \\
 \delta\phi^{\textrm{coor}}(x)&=&\int d^n x' K^{\phi(x)}{}_\lambda(x,x')\delta\epsilon^\lambda(x')\,,
\end{eqnarray}
with the symmetry generators defined as
\begin{eqnarray}\label{generators}
 K^{g_{\mu\nu}(x)}{}_\lambda(x,x')&=&- g_{\mu\nu,\lambda}(x)\delta(x,x')-2g_{\lambda(\nu}(x)\partial_{\mu)}\delta(x,x'), \\
 K^{\phi(x)}{}_\lambda(x,x')&=&-\partial_\lambda\phi(x)\delta(x,x').
\end{eqnarray}

In a gauge theory, the functional measure in the generating functional is built out of infinitesimal field configurations defined up to a gauge transformation, and a gauge condition is required to ensure integrating only over physical field configurations. In this regard, it turns out important to distinguish between those field redefinitions induced by a gauge transformation from those which are not.
To distinguish between the two types of field displacements consider the following decomposition of an arbitrary field displacement into parallel and perpendicular components to the gauge vectors,
\be
\delta \Phi^i = \delta_{||}  \Phi^i + \delta_{\perp} \Phi^i =  K^{i}_{\alpha} d\epsilon^\alpha +  \delta_{\perp} \Phi^i .
\ee
A gauge-fixing condition, $\chi^{\alpha}[\Phi^i] = 0$, then introduces a set of ``trajectories'' or ``orbits'' in the space of fields, called the ``gauge orbits'', with each orbit intersecting the surface $\mathcal{S}$ defined through the gauge condition at a unique point. Given this geometrical approach to the gauge--fixing of the theory, each gauge orbit can be parametrised using a coordinate system with coordinates $\chi[\Phi]^A, \xi[\Phi]^A$, with the first ones running along the orbit, while the latter ones parametrising the ``constant-orbit surface'' $\mathcal{S}$. The latter, are by construction gauge invariant, while the former describe different points along a gauge orbit. 

One can define the line element in field space using the above decomposition for the infinitesimal field variations as
\begin{align}
ds^2 = \mathfrak{g}_{ij}d\Phi^i d\Phi^j & = \mathfrak{g}_{ij}\left(d\Phi_{||}^i + d \Phi_{\perp}^i \right) \cdot \left(d \Phi_{||}^j + d \Phi_{\perp}^j \right)  = \mathfrak{g}_{ij}d \Phi^{i}_{||}d \Phi^{j}_{||} + \mathfrak{g}_{ij}d \Phi^{i}_{\perp} d \Phi^{j}_{\perp}\nonumber \\
& = \gamma_{\alpha \beta}d \epsilon^\alpha d \epsilon^\beta + \mathfrak{g}_{ij}^{\perp}\omega^{i}_{\perp} \omega^{j}_{\perp},
\end{align}
with the last line following from orthogonality. $\mathfrak{g}_{ij}$ is the metric in field space, and $d \Phi^i_{\perp} = P^{i}{}_{j} d \Phi^j$, $d \Phi^i_{||} =K^{i}_{\alpha} d \epsilon^\alpha$, with $P^{i}{}_{j}$ the projection operator onto the constant-gauge hypersurface $\mathcal{S}$ satisfying $P^{i}_{j}K^{j}_{\alpha} = 0$. The perpendicular part of the metric can be extracted using the projection operator as $\mathfrak{g}_{ij}^{\perp} = P^{m}_{i} P^{n}_{j} \mathfrak{g}_{mn}$.
One can formally then show that the induced metric on $\mathcal{S}$ is given by $h_{AB} = \Phi^{n}_{,A} \Phi^{m}_{,B} \mathfrak{g}_{nm}^{\perp}$, and the requirement that $\mathcal{S}$ is a gauge-invariant (constant gauge-orbit manifold), i.e $\delta h^{AB}/\delta \theta^\alpha = 0$, implies that the field-space metric satisfies the Killing equation,
\begin{equation}
\mathfrak{g}_{ij,k}K^k_\alpha+2K^k_{\alpha,(i}\mathfrak{g}_{j)k}=0.
\end{equation}
The latter equation is essentially a differential equation for $\mathfrak{g}_{ij}$ with the only theory input to it being the generators $K_{i \mu}$, and any consistent choice of field--space metric should satisfy it. 

The further requirements of ultralocality, that is, no dependence on derivatives of the original fields to avoid extra derivatives in the connection, and of diagonality of the field-space metric bring us to the choice of $\mathfrak{g}_{ij}$. DeWitt \cite{deW} suggested that the unique metric which does not introduce new dimensionful parameters, is given by
\begin{align}
& \mathfrak{g}_{g_{\mu \nu}(x) g_{\rho \sigma}(x')} = \frac{1}{\kappa^2} \sqrt{g(x)}\cdot \left( g^{\mu (\rho}g^{\sigma) \nu}  - \frac{1}{2}g^{\mu \nu}g^{\rho \sigma} \right)\delta(x,x'),\label{ggg}\\
& \mathfrak{g}_{\phi(x) \phi(x')} = \sqrt{g(x)}\delta(x,x').
\end{align}
Note that the constant coefficient $\kappa^{-2}$ in \eqref{ggg} is needed for two reasons: it is contained in the highest derivative of the metric in the starting action \eqref{action:0}, and is necessary to fix the dimensionality of the effective action. 
Given an expression for $\mathfrak{g}_{ij}$, the connection in field space can be calculated in the standard way as
\be
\Gamma^{k}_{ij} =  \frac{1}{2}  \mathfrak{g}^{\kappa l}(2 \partial_{(j} \mathfrak{g}_{i)l} - \partial_l  \mathfrak{g}_{ij}),
\ee
where it is understood that the partial derivatives here play the role of functional derivatives with respect to the corresponding field. The presence of the connection terms provide new interactions in the effective action, revealing novel features as we will see later. 

We now ask the following question: Is the particular choice of field--space metric consistent with Galileon symmetry\footnote{The authors are thankful to Tim Morris for interesting discussions around this point.}? The metric is by construction compatible with the general coordinate transformation, but the theory enjoys an extra global symmetry. Looking at the Galileon transformation (\ref{Galileon-symmetry}), the corresponding generators follow as \cite{Hinterbichler:2015pqa}
\begin{align}
&C \cdot \phi = 1, \; \; B^{\mu} \cdot \phi = x^\mu \label{Galilean-generators},
\end{align}
with the first corresponding to the standard shift-symmetry (with the generator being simply the unity operator), and the second to the generator of the coordinate-dependent piece of the Galileon transformation. One notices that both generators are independent of the scalar field $\phi$. Now, choosing $k = \phi$ in the Killing equation one finds that
\[
 \mathfrak{g}_{ij,\phi}K^{\phi}_{\alpha} + 2 K^{\phi}_{\alpha, (i} \mathfrak{g}_{j)\phi} = 0,
\]
but in view of the constant generators for the Galilean symmetry (\ref{Galilean-generators}) we have that $K^{\phi}_{\alpha, i} = 0$. The Killing equation then yields the constraint
\[
 \mathfrak{g}_{ij,\phi} = 0,
\]
which suggests that the super-metric is $\phi$-independent. It is true that the Killing equation as defined above can be in principle generalised to include source terms parallel to the Killing vectors, and one could always think of more general super-metrics which would satisfy it -- however, this would only lead to a more complicated form for the super-metric, without affecting the structure of the Killing vectors. We conclude that no extra Killing vectors need to be introduced for the Galileon symmetry to be respected by our choice for the field--space metric.

For the evaluation of the effective action at 1--loop, we will need the action expanded up to second-order in field fluctuations. In the covariant setting we are considering here, a crucial point is that the functional derivatives used to evaluate the second--order action are promoted to covariant functional derivatives in field space associated with $ \mathfrak{g}_{ij}$, and we denote them as $\nabla_i$. The quadratic action then reads as
\begin{align}
S_{\text{quad}} = \frac{1}{2} \lim_{\alpha \to 0} \eta^i \eta^j \left( \nabla_i \nabla_j S + \frac{1}{2\alpha} K^{\alpha}_{i} K_{j \alpha} \right), \label{eq:S_quad}
\end{align}
with the metric and scalar field fluctuations $\eta^{i} = \{h_{\mu \nu}, \psi \}$. The second piece on the right-hand side of above equation corresponds to the gauge-fixing part and $\alpha$ is the gauge-parameter which has to be evaluated to zero at the end of the calculation, a choice corresponding to the gauge--invariant result of the DeWitt gauge. 
The covariant derivatives are evaluated in the usual way, 
\be\label{eq:cov-action}
\nabla_i \nabla_j S = \partial_i \partial_j S  - \Gamma^{k}_{ij} \partial_{k} S,
\ee
where $ \Gamma^{\kappa}_{ij}$ is the Christoffel connection built out of the field-space metric. Notice that, when evaluated on--shell ($\partial_{k} S = 0$) the effective action reduces to the standard one. This is intimately related to the fact that observable (on--shell) quantities should not depend on the choice of background and gauge. 
With the super-metric in hand, we can calculate the corresponding Christoffel symbols in the usual way. The one associated purely with the gravitational sector reads as
\begin{align}\label{eq:Gammas}
 \Gamma^{g_{\lambda\tau}(x)}_{g_{\mu\nu}(x')g_{\rho\sigma}(x'')}\equiv\Gamma^{\mu \nu \rho \sigma}_{\lambda \tau} = - \delta^{(\mu}_{(\lambda}g^{\nu) (\rho}\delta^{\sigma)}_{\tau)} +& \frac{1}{4}g^{\mu \nu} \delta^{\rho}_{(\lambda} \delta^{\sigma}_{\tau)} + \frac{1}{4} g^{\rho \sigma} \delta^{\mu}_{(\lambda} \delta^{\nu}_{\tau)}  +\nonumber\\
 &+\frac{1}{2(n-2)}\left(g_{\lambda\tau}g^{\mu(\rho}g^{\sigma)\nu}-\frac{1}{2}g_{\lambda\tau}g^{\mu\nu}g^{\rho\sigma}\right)\,,
\end{align}
with $n$ denoting the number of spacetime dimensions, while for the rest we have
\begin{align}\label{eq:Gammas2}
& \Gamma^{\phi(x)}_{\phi(x')\phi(x'')}=\Gamma^{g_{\mu\nu}(x)}_{\phi(x')g_{\rho\sigma}(x'')}=\Gamma^{\phi(x)}_{g_{\mu\nu}(x')g_{\rho\sigma}(x'')}=0\,,\nonumber\\
& \Gamma^{g_{\mu\nu}(x)}_{\phi(x')\phi(x'')}=\frac{\kappa^2}{2(n-2)}g_{\mu\nu}(x)\delta(x,x')\delta(x'',x')\,,\nonumber\\
& \Gamma^{\phi(x)}_{\phi(x')g_{\mu\nu}(x'')}=\frac{1}{4}g^{\mu\nu}(x)\delta(x,x')\delta(x'',x').
\end{align}
When computing the quadratic action, we assume a background--field expansion, $\Phi^i = \bar \Phi^i  + \eta^i $, with $\bar \Phi^i$ denoting a background and $\eta^i$ a fluctuating field respectively,\footnote{Notice that the metric fluctuation is defined to have mass dimensions one.}
\begin{align}
& g_{\mu \nu}(x) = \bar{g}_{\mu \nu} + \kappa h_{\mu \nu}, \; \; \label{Field-expansion}
 \phi(x) = \bar{\phi} + \psi.
\end{align}
Since we are dealing with a background-independent formalism we will set $\bar{g}_{\mu \nu} = \delta_{\mu \nu}$ (in Euclidean signature) for simplicity. In what follows, for convenience, we will also drop the overbars from background quantities.
In view of (\ref{Field-expansion}) and the symmetry generators \eqref{generators}, the gauge-fixing condition $\chi_\nu$ becomes
\begin{align}
\chi_\nu=K_{i \nu} \eta^{i} = \frac{2}{\kappa}\left( \partial^\mu h_{\mu \nu} - \frac{1}{2}\partial_\nu h\right) - \omega \partial_{\nu} \phi \psi,
\end{align}
with $\omega$ a book-keeping parameter to keep track of the terms coming from the scalar-field piece in the gauge-fixing condition $\chi_\nu = 0$.

For the connection-dependent part of the effective action we will need the expression for the energy--momentum tensor and equation of motion for the scalar field, $T^{\mu \nu}$ and $\mathcal{E}_{\phi}$, which have been calculated in (\ref{EOM}) and (\ref{EMT}) respectively. Collecting all terms together from (\ref{eq:S_quad}) we then have 
\begin{align}
 & \frac{1}{2} \eta^i \eta^j \left( \nabla_i \nabla_j S + \frac{1}{2\alpha} K^{\alpha}_{i} K_{j \alpha} \right)   = 
 \frac{1}{2} \psi \cdot \left[ \partial_{\phi \phi}^{2} S\right] \cdot \psi  +   \frac{1}{2} h_{\alpha \beta} \cdot \left[  \partial_{gg}^{2} S\right]^{\alpha \beta \gamma \delta} \cdot h_{\gamma \delta} \nonumber \\
& +  \frac{1}{2} \int d^{4}x  \Bigg\{ -\gamma\kappa^2 h_{\mu \nu} \cdot h_{\rho \sigma} \cdot  \Gamma^{\mu \nu \rho \sigma}_{\lambda \tau} T^{\lambda \tau} 
- \frac{\gamma}{4}\psi^2\cdot \kappa^2 g_{\mu \nu} T^{\mu \nu} 
- \frac{\gamma}{2} \psi \cdot \kappa h_{\mu \nu}  \cdot g^{\mu \nu} \mathcal{E}_{\phi}   \nonumber \\
 & \hspace{3cm}+ \frac{1}{ 2\alpha}\left[ \frac{2}{\kappa}\left( \partial^{\mu} h_{\mu \nu}  - \frac{1}{2} \partial_{\nu} h\right) -   \omega\, \psi \cdot  \phi_\nu  \right]^2
 \Bigg\}.
\end{align}
Evaluating the field--space Christoffel symbols, the energy-momentum tensor and equation of motion for the scalar, we arrive at the more explicit expression 
\begin{align}
  \eta^i \eta^j \left( \nabla_i \nabla_j S  + \frac{1}{2\alpha} K^{\alpha}_{i} K_{j \alpha} \right) & = \frac{1}{2} \psi \cdot \left[ \partial_{\phi \phi}^{2} S\right] \cdot \psi  +   \frac{1}{2} h_{\alpha \beta} \cdot \left[  \partial_{gg}^{2} S\right]^{\alpha \beta \gamma \delta} \cdot h_{\gamma \delta}  \nonumber \\
& \hspace{-4.5cm}+  \int d^{4}x \Bigg[ \gamma \kappa^2\left( \frac{1}{16} \phi^{\alpha}\phi_{\alpha} h_{\rho \sigma} h^{\rho \sigma} - \frac{1}{32} \phi^{\alpha}\phi_{\alpha} h^2  - \frac{1}{4} \phi^{\sigma} \phi^{\mu} h_{\mu}{}^{\rho} h_{\rho \sigma} + \frac{1}{8} \phi^\rho \phi^\sigma h h_{\rho \sigma} \right) - \frac{\gamma \kappa^2}{16} \phi^{\alpha}\phi_{\alpha} \psi^2 \nonumber \\
&\hspace{-4cm}+ \frac{\gamma}{4} \kappa\psi h \left[ \Box \phi(1+B)  - M^{-3}\phi_{\alpha \kappa}\phi^{\alpha \kappa}\right]+ \frac{1}{ 4\alpha}\left[ \frac{2}{\kappa}\left( \partial^{\mu} h_{\mu \nu}  - \frac{1}{2} \partial_{\nu} h\right) -   \omega\, \psi \cdot  \phi_\nu  \right]^2\nonumber\\
&\hspace{-3.5cm}-\frac{\gamma}{2}\Lambda\,\frac{n-4}{n-2}\cdot\left(\frac{h^2}{2}
 -h^{\mu\nu}h_{\mu\nu}\right)-\gamma\frac{ n \Lambda}{2n-4}\psi^2 \Bigg], \label{Explicit-connection}
\end{align}
where in the last expression we made explicit the limit $n\rightarrow4$ in the second and third line, but not in the last line; this choice is in order to ease the notation, but also to show how the effective masses of the propagators are modified as a matter of the spacetime dimensionality and cosmological constant respectively. We have checked that taking {\em a priori} the limit to four dimensions for the terms coming from the second and third line will not affect the final results.
Notice that the second-order pieces arising from the standard variation of the bare action (i.e with respect to $\partial_i$) will be presented later. Notice also the two book--keeping parameters: The parameter $\gamma$ which traces the terms coming from the field--space connection and the parameter $\omega$ coming from the gauge--fixing part of the scalar field respectively. Both of them should be set equal to one at the end of the calculation to derive the field--re-parametrisation invariant result. 

Most importantly, one notices the mass--type interaction for the scalar, $\sim \gamma \Lambda \psi^2$. Its origin is purely geometrical, due to the presence of the connection terms in the evaluation of the second--order action, and as we will show later it will give an effective mass to the scalar propagator, which will play a crucial role in revealing a new operator structure in the effective action. Obviously, it vanishes for a flat field--space connection, $\gamma = 0$.

\section{1--loop effective action and the structure of divergences} \label{sec:Calculation}
We are now in the position to introduce the effective action and start discussing its evaluation at 1--loop. Given a bare action $S$, the associated effective action $\Gamma$ is defined as
\begin{align}
\Gamma = -\ln \int [d\eta]e^{- S_{\text{quad}}[\bar{\Phi}; \eta] },  \label{action:effective0}
\end{align}
with the Gaussian piece of the bare action $S$ defined in (\ref{eq:S_quad}). The quadratic action then organises itself in powers of the background field as 
\begin{equation}
S_{\text{quad}} = S[\bar{\Phi}^0; \eta] + S[\bar{\Phi}; \eta] + S[\bar{\Phi}^2; \eta] + \ldots \equiv S_0 + \delta S.
\end{equation}
The zeroth--order piece, $S_0$, allows to read off the propagators of the different fields, which we will perform in momentum space later on. Our approach will be perturbative, which means that, we will be treating $\delta S$ as a small interaction term. Truncating to different powers of the background field and evaluating the trace over field fluctuations, will reveal the renormalisation of the different operators of the original theory. On the same time, the result of this computation will also suggest the new (higher--order) terms generated by quantum corrections, and which would have to be in principle included in the original action for consistency under renormalisation. The effective action can be then perturbatively evaluated as
\be
\Gamma \simeq - \ln \int [d \eta] e^{-S_0} \left( 1 - \delta S + \frac{1}{2} \delta S^2\right) = - \ln \left( 1 +  \Braket{\delta S} - \frac{1}{2} \Braket{\delta S^2}  \right),
\ee
which can be further expanded using $\ln(1+x) \simeq x - x^2/2$. As explained earlier, we will evaluate the trace over field fluctuations perturbatively truncating up to second--order in the background field, i.e 
\be
\delta S \equiv  S_{1} + S_{2},
\ee
with the index denoting the order in the background scalar. This means that we will be unable to capture the renormalisation of the cubic Galileon term itself, however, its presence will have a non---trivial effect to lower--order interactions. Under these assumptions, the evaluation of the effective action then boils down to calculating  
\be
\Braket{S_2} - \frac{1}{2}\Braket{S_{1}^2} \equiv \Braket{S_2(x,x)}- \frac{1}{2}\Braket{S_{1}(x) S_{1}(y)} \label{Pairings0}.
\ee
Above pairings correspond to infinite trace integrals, and their divergent piece will define the pole--part of the effective action. To isolate the divergent piece we will use the scheme of dimensional regularisation which manifestly preserves the gauge symmetries of the theory. 

We now proceed with presenting the expression for the quadratic action explicitly, followed by the evaluation of the 1--loop effective action. 
Let us first introduce the following convenient notation
\be
\phi_{\alpha}(x) \equiv \partial_{\alpha} \phi(x), \; \; \psi_{\alpha}(x) \equiv \partial_{\alpha} \psi(x). 
\ee
The explicit intermediate steps of the calculation are presented in the appendix. 
For the zeroth--, first-- and second--order quadratic action in the background scalar we have,
\begin{align}
S_0&=\int d^d x \cdot \left\{\frac{1}{2}\delta^{\mu\nu}\psi_\mu\psi_\nu-\frac{1}{2}h^{\mu\nu}\Box h_{\mu\nu}+\frac{1}{4}h \Box h+\left(\frac{1}{\alpha\kappa^2}-1\right)
 \left(\partial^\mu h_{\mu\nu}-\frac{1}{2}\partial_\nu h\right)^2+\right.\nonumber \\
 &\left.\hspace{1cm}{+\Lambda\left(\frac{h^2}{2}
 -h^{\mu\nu}h_{\mu\nu}\right)\left[1-\frac{\gamma}{2}\left(\frac{n-4}{n-2}\right)\right]}{
 -\gamma\Lambda\frac{n}{2n-4}\psi^2}\right\}, \label{eq:S_0}
\end{align}
\begin{align}
S_1 &=\! \int  d^d x \cdot \left\{\frac{1}{M^3}\left(\frac{1}{2}\Box\phi\psi_\mu\psi^\mu+\phi_\mu\psi^\mu\Box\psi\right)
+\frac{\kappa}{2} \phi_\mu h\psi^\mu -\kappa h^{\mu\nu}\psi_\mu\phi_\nu
 +\frac{\kappa\gamma}{4}h\psi\Box\phi-\right.\nonumber\\
&\left.\hspace{3cm}-\frac{\omega}{\alpha\kappa}\left(\partial^\lambda h_{\lambda\nu}-\frac{1}{2}\partial_\nu h\right)\phi^\nu~\psi\right\}, \label{S1:explicit}
\end{align}
\begin{align}
 S_2&= \kappa^2\int  d^d x \cdot \left\{\frac{1}{2}X\left(\frac{h^2}{4}-\frac{1}{2}h_{\mu\nu}h^{\mu\nu}\right)-\frac{1}{4}hh^{\mu\nu}\phi_\mu\phi_\nu+\frac{1}{2}h^\mu{}_\alpha
 h^{\alpha\nu}\phi_\mu\phi_\nu+\right.\nonumber\\
 &+\frac{1}{\kappa M^3}\left[\frac{h}{2}\Big(\Box\phi \phi_{\mu}\psi^\mu+X\Box\psi\Big)-h^{\mu\nu}\psi_\mu\phi_\nu\Box\phi +\phi_\beta~\psi^\beta\left(-h^{\mu\nu}\phi_{\mu\nu}-\frac{1}{2}\phi^{\rho}
\left(2\partial_\mu h^{\mu}{}_{\rho}-\partial_\rho h\right)\right)\right.\nonumber \\
&\left.\hspace{3cm}-\frac{1}{2}h^{\mu\nu}\phi_\mu\phi_\nu\Box\psi+X\left(-h^{\mu\nu}\psi_{\mu\nu}-\frac{1}{2}\psi^{\rho}
\left(2\partial_\mu h^{\mu}{}_{\rho}-\partial_\rho h\right)\right)\right]+\nonumber \\
&\left.+\gamma\left[-\frac{1}{16}\psi^2\phi_\mu\phi^\mu
+\frac{1}{16} h_{\mu\nu} h^{\mu\nu} \phi_{\lambda} \phi^{\lambda} - \frac{1}{32} h^{\mu}{}_{\mu} h^{\nu}{}_{\nu} \phi_{\lambda} \phi^{\lambda}-\frac{1}{4}h_{\lambda}{}^{\nu} h_{\mu\nu} \phi^{\lambda} \phi^{\mu} +\frac{1}{8} h_{\lambda\mu} h^{\nu}{}_{\nu} \phi^{\lambda} 
\phi^{\mu}\right]+\right.\nonumber\\
&\left.\hspace{3cm}+\frac{\gamma}{\kappa M^3}\frac{h}{4}\psi(\Box\phi\Box\phi-\phi^{\alpha\mu}\phi_{\alpha\mu})+\frac{\omega^2}{4\alpha\kappa^2}\phi^\mu\phi_\mu~\psi^2\right\}. \label{S2:explicit}
\end{align}


From $S_0$ we can read off the momentum--space propagators associated with the graviton and scalar as\footnote{We will be using the convention: $\mathcal{G}(x,y) = \int \frac{d^n p}{(2 \pi)^n} \mathcal{G}(p) \cdot e^{i p (x-y)}$.}
\begin{align}
G_{\alpha\beta\gamma\delta}(p)=&\frac{\delta_{\alpha\gamma}\delta_{\beta\delta}+\delta_{\alpha\delta}\delta_{\beta\gamma}
 -\frac{2}{n-2}\delta_{\alpha\beta}\delta_{\gamma\delta}}{2(p^2{-2\lambda})}+\nonumber \\
&\hspace{-1cm}+ (\alpha-1)\frac{\delta_{\alpha\gamma}p_{\beta}p_{\delta}
 +\delta_{\alpha\delta}p_{\beta}p_{\gamma}+p_{\alpha}p_{\gamma}\delta_{\beta\delta}+p_{\alpha}p_{\delta}\delta_{\beta\gamma}}
 {2 (p^2{-2\lambda})(p^2{-2\alpha\kappa^2\lambda})},  
\end{align}
\begin{align}
G(p)=\frac{1}{p^2+m_\Lambda^2}, \; \; m_\Lambda^2 = \gamma \cdot \frac{n\Lambda}{2-n}, \label{eq:Scalar-prop}
\end{align}
with the definition $\lambda \equiv\Lambda+\gamma\Lambda\left(\frac{n-4}{4-2n}\right)$. 

An important point is in order here. Since the original theory is shift-symmetric, one would have expected the scalar propagator to be massless. However, the latter acquires a mass $\sim \Lambda$, due to the mass--type interaction in (\ref{eq:S_0}), which' origin is identified in the term $\Gamma^{g}_{\phi \phi} \partial_{g} S$ of the connection--dependent piece of (\ref{eq:cov-action}) (see also \eqref{eq:Gammas2}), and is absent for a flat field-space metric ($\gamma = 0$). As we will see shortly, it will have an important impact on the 1--loop structure of the effective action. We should emphasise the gravitational origin of this interaction: in a scalar theory without gravity, the vacuum can be always removed as an unphysical contribution, but this is no longer true as long as gravity is present. 

\subsection{Trace integrals}
We will be working in momentum space, and it is instructive to start by illustrating with an example the structure of the integrals we will be 
calculating. We will denote with $\Phi_i(x)$ an arbitrary field with effective mass $m_i$ and propagator $\mathcal{G}_i(x,x')$, 
and $F, B$ some abstract functions of background fields. The typical integral involved in the calculations will then schematically contain $(j+k)$ 
derivatives of the propagators, finally reading
\begin{align}\label{integrals}
&  F(x) \cdot B(x') \cdot \Braket{ \Phi_1(x) \Phi_1(x') } \cdot \partial^{(j)} \partial^{' (k)} \Braket{\Phi_2(x) \Phi_2(x') }=\nonumber \\
&  = \dint d^nx\, d^nx'\cdot F(x) \cdot B(x')\cdot \mathcal{G}_1(x,x') \cdot\partial^{(j)} \partial^{' (k)} \mathcal{G}_2(x,x') \nonumber \\
& =   \dint d^nx\, d^nx'\cdot F(x) \!\!\int\! \!\frac{d^nq}{(2\pi)^n}\,\! B(q) e^{iqx'} \! \!\int\!\! \frac{d^nl}{(2\pi)^n} \,\!
e^{il(x-x')}\mathcal{G}_1(l) \!\int\! \! \frac{d^ns}{(2\pi)^n} e^{is(x-x')} (i s)^j \!\cdot\! (-i s)^k \mathcal{G}_2(s)\nonumber \\
& =   \dint d^nx\, d^nx'\cdot F(x)\int  \frac{d^nq}{(2\pi)^n} B(q) e^{iqx}  \int  \frac{d^ns}{(2\pi)^n} (-1)^k (i s)^{j+k}
\cdot \mathcal{G}_1(q-s)\cdot \mathcal{G}_2(s). 
\end{align}
From here on, a possible strategy would be to calculate the integrals using the method of the Feynman parameters to combine the product of 
Green functions, and any available scheme to regularise the integrals. This procedure is formally the most straightforward for 
calculating such integrals, and it also provides a way to evaluate exactly the finite part of the effective action, however, it requires very demanding calculations. For this reason, and since we are here only interested in the UV behaviour of the integrals, 
below we will follow another approach: we will be first evaluating the integrand asymptotically for large (internal) momenta and will keep 
only the logarithmically divergent piece to track the pole part. Let us be more specific. The last integral 
appearing in the last line of \eqref{integrals} can be explicitly written in momentum space as
\be
I = \int  \frac{d^ns}{(2\pi)^n} \frac{s_{\mu_1} s_{\mu_2} \cdots s_{\mu_{j+k}}}{  \left[ (q_{\mu}-s_{\mu})^2  + m^2_1 \right]\cdot\left(s^2 + m^2_2 \right)} = 
\int \frac{d^ns}{(2\pi)^n}  \frac{s^{j+k} \cdot \hat{s}_{\mu_1} \hat{s}_{\mu_2} \cdots \hat{s}_{\mu_{j+k}}}{ \left[ s^2 + q^2 + 2s q \cdot \theta  + m^2_1 \right]\cdot \left(s^2 + m^2_2 \right)} 
\label{Main-integral}, 
\ee
with $s^2 \equiv \delta^{\mu \nu} s_{\mu} s_{\nu}$, $s \equiv \left|\sqrt{s^2}\right|$, and $\theta \equiv \hat{s}_{\mu} \hat{q}^{\mu}$ (being 
$\hat{s}^{\mu}$ and $\hat{q}^{\mu}$ the unit vectors along the $\mu-$direction). It turns out convenient to expand the integrand
for large momenta $s$, truncating the expansion at the logarithmic pole $\sim s^{-n}$.
The product of the unit vectors $\hat{s}_{\mu_i}$ appearing in the numerator\footnote{In this regard, it becomes crucial to take into account also the 
further $s_\mu$ factors coming from the function $\theta$ after the large momenta expansion.} of 
\eqref{Main-integral}
can be sensibly simplified: any product involving an odd number of powers of $s$ vanishes because of symmetric integration, 
while an even power can be replaced by the even terms
\begin{align}\label{replace}
& \hat{s}_{\mu} \hat{s}_{\nu} = \frac{1}{n} \delta_{\mu \nu},\qquad \hat{s}_{\mu} \hat{s}_{\nu} \hat{s}_{\rho} \hat{s}_{\sigma} = \frac{1}{n (n+2)} \left( \delta_{\mu \nu} \delta_{\rho \sigma} + \delta_{\mu \rho} \delta_{\nu \sigma} 
+ \delta_{\mu \sigma} \delta_{\nu \rho} \right),\nonumber\\
& \hat{s}_{\mu} \hat{s}_{\nu} \hat{s}_{\rho} \hat{s}_{\sigma} \hat{s}_\chi \hat{s}_\xi  = \frac{1}{n (n+2) (n+4)} \left( \delta_{\mu \nu} \delta_{\rho \sigma} \delta_{\chi\xi}+ 
\textrm{permutations} \right),
\end{align}
and similar expressions hold for higher powers. Given such prescriptions every integral like \eqref{Main-integral} will then be re-written in terms of the 
logarithmically divergent integral,
\begin{equation}
a_L=\int   \frac{d^ns}{(2\pi)^n} \frac{1}{s^n}\;\;\stackrel{n\rightarrow4}{\xrightarrow{\hspace{1cm}}}\;\;a_L=-\frac{1}{8\pi^2(n-4)}\,,
\end{equation}
where the limit to four dimensions is here understood.

\subsection{Pole contribution from the tadpole--like term}

\begin{figure}[!htpb]
\centering
\includegraphics[width=\textwidth]{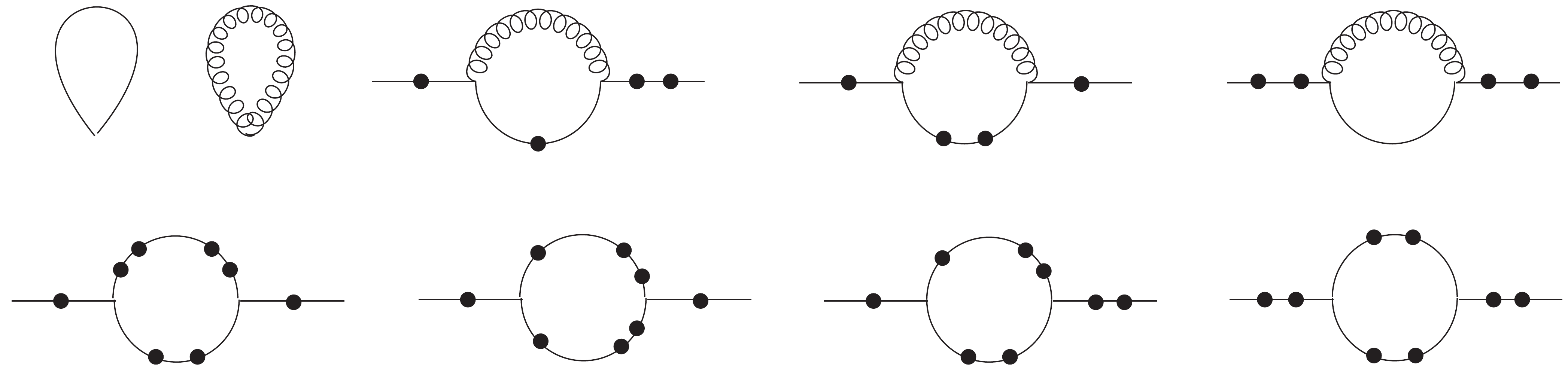}
\caption{Feynman--diagram representation of the contributing terms at 1--loop level. Solid lines represent scalar, while curly ones graviton propagators respectively. Solid external lines denote background scalar fields, and full dots derivatives acting on the corresponding propagators/background fields. The first two diagrams on the upper line correspond to the contribution of $\Braket{S_2(x,x)}$, with the first one coming purely from the field--space connection/gauge sector. The rest three diagrams on the upper line (``grandma'' diagrams) correspond to graviton--scalar interactions in $\Braket{S_{1}(x) S_{1}(x')}$, while the ones in the second line to the scalar--scalar contribution of the same term. }
\label{Granma}
\end{figure}

It turns out to be convenient to split the evaluation of the two pieces appearing in the 1--loop effective action, starting from the tadpole-like term 
$\langle S_2(x,x)\rangle$, which is the term involving uniquely the coincidence limit of the Green functions. 
Discarding terms involving an odd number of fields in view of Wick's theorem\footnote{Note that for this reason, both the second 
and the third line of \eqref{S2:explicit}, being proportional to the term $h(x)\cdot\psi(x)$, 
will not contribute to $\langle S_2(x,x)\rangle$.}, 
\be
\langle h_{\mu\nu}(x) \psi(x')\rangle=0,
\ee
we get a purely scalar and a purely gravitational contribution of a tadpole--type,
\begin{align}
 \langle S_2(x,x)\rangle&=\int d^nx\cdot \kappa^2\Bigg[\left(\frac{1}{8}-\frac{\gamma}{16}\right) X \langle h^2\rangle
 +\left(\frac{\gamma}{8}-\frac{1}{4}\right) X \langle h_{\mu\nu}h^{\mu\nu}\rangle
+ \left(\frac{\gamma}{8}-\frac{1}{4}\right) \phi^\mu\phi^\nu \langle h h_{\mu\nu}\rangle+\nn
&+\left(\frac{1}{2}-\frac{\gamma}{4}\right) \phi^\mu\phi^\nu \langle h_{\mu\lambda}h^{\lambda}{}_\nu\rangle
+\left(\frac{\omega^2}{2\alpha\kappa^2}-\frac{\gamma}{8} \right)X \langle\psi^2\rangle\Bigg]=\nn
&=\int d^n x \cdot \kappa^2\left\{\left[\left(\frac{1}{8}-\frac{\gamma}{16}\right) X ~\delta^{\mu\nu}\delta^{\lambda\sigma}
+\left(\frac{\gamma}{8}-\frac{1}{4}\right) X~\delta^{\mu\lambda}\delta^{\nu\sigma}
+ \left(\frac{\gamma}{8}-\frac{1}{4}\right) \phi^\mu\phi^\nu \delta^{\lambda\sigma}\right.\right.\nn
&\left.\left.+\left(\frac{1}{2}-\frac{\gamma}{4}\right)\phi^\mu\phi^\sigma\delta^{\nu\lambda}\right] G_{\mu\nu\lambda\sigma}(x,x)\!
+\!\left(\frac{\omega^2}{2\alpha\kappa^2}-\frac{\gamma}{8} \right)\!X G(x,x)\right\} \equiv \langle S_2\rangle_{\textrm{grav}}+\langle S_2\rangle_{\textrm{scal}}\,.
\end{align}
The pole parts of the two contributions can be extracted with the method described above, that is, moving first to momentum space and then expanding the integrand in (inverse) powers of internal momenta $\sim p^{-1}$, and in the limit of large momenta, truncating the power series at the logarithmically divergent term $p^{-n}$. The pole part of the integral can be then extracted using the gauge--covariant technique of dimensional regularisation. Considering the limit to four dimensions, the contributions to the pole terms coming from 
$\langle S_2\rangle_{\textrm{grav}}$ and $\langle S_2\rangle_{\textrm{scal}}$ will respectively read
\begin{align}\label{graviton}
 \langle S_2\rangle_{\textrm{grav}}^{\textrm{pole}}&= \int d^n x \cdot \kappa^2\left[\left(\frac{1}{8}-\frac{\gamma}{16}\right) X ~\delta^{\mu\nu}\delta^{\lambda\sigma}
+\left(\frac{\gamma}{8}-\frac{1}{4}\right) X~\delta^{\mu\lambda}\delta^{\nu\sigma}
+ \left(\frac{\gamma}{8}-\frac{1}{4}\right) \phi^\mu\phi^\nu \delta^{\lambda\sigma}\right.\nn
&\left.+\left(\frac{1}{2}-\frac{\gamma}{4}\right)\phi^\mu\phi^\sigma\delta^{\nu\lambda}\right] G_{\mu\nu\lambda\sigma}(x,x)=0\quad
\textrm{(modulo finite terms)},
 \end{align}
 \begin{align}\label{scalar}
\langle S_2\rangle_{\textrm{scal}}^{\textrm{pole}}&=\int d^nx\left(\frac{\omega^2}{2\alpha}-\frac{\gamma\kappa^2}{8} \right)X G(x,x)=
 \int d^nx\left(\frac{\omega^2}{2\alpha}-\frac{\gamma\kappa^2}{8} \right)X \int \frac{d^np}{(2\pi)^n}\frac{1}{p^2+m^2_\Lambda}=\nonumber\\
 &=-a_L m^2_\Lambda\left(\frac{\omega^2}{2\alpha}
 -\frac{\gamma\kappa^2}{8} \right)\int d^nx X\,.
 \end{align}
The corresponding representation in terms of Feynman diagrams can be seen in the first two diagrams of Fig. \ref{Granma}.
It is interesting to note here that the first term (purely gravitational contribution), $\langle S_2\rangle_{\textrm{grav}}$, contributes only a finite term, 
thus it does not contribute to the divergent piece of the effective action. Moreover, we have further checked that, this is true 
not only in the limiting case of $n\rightarrow4$, but for any number of dimensions. Notice also that, the scalar contribution includes only terms coming from the gauge--fixing part associated with the scalar ($\sim \omega/\alpha$) and the connection ($\sim \gamma$). The latter contribution is crucial as it will cancel a similar term from the pairing $\langle S_1(x)S_1(x')\rangle$ to be presented in the next subsection, ensuring the finiteness of the result in the limit $\alpha \to 0$.

\subsection{Pole contribution from scalar--scalar and graviton--scalar interactions}

We are now interested in computing the contribution coming from the scalar--scalar and graviton--scalar interactions, stemming from $\langle S_1(x)S_1(x')\rangle$, which corresponds to the most cumbersome part of the calculation. 
Using the expression \eqref{S1:explicit} for $S_1(x)$ and re-arranging the terms, one finds
\begin{align}
S_1&=\int d^nx \left[\frac{1}{M^3}\left(\frac{1}{2}\Box\phi\partial_\mu\psi\partial^\mu\psi+\partial_\mu\phi\partial^\mu\psi\Box\psi\right)+\frac{\kappa h}{2}\partial_\mu\phi\partial^\mu\psi-\kappa h^{\mu\nu}\partial_\mu\psi\partial_\nu\phi
 +\frac{\gamma}{4}\kappa h\psi\Box\phi-\right.\nn
&\left.\hspace{3cm}-\frac{\omega}{\alpha\kappa}\left(\partial^\lambda h_{\lambda\nu}-\frac{1}{2}\partial_\nu h\right)\partial^\nu\phi~\psi\right]=\nn
&=\int d^nx \left[\frac{1}{M^3}\left(\frac{1}{2}\Box\phi\partial_\mu\psi\partial^\mu\psi+\partial_\mu\phi\partial^\mu\psi\Box\psi\right)
+P^{\alpha\beta\mu}h_{\alpha\beta}\partial_\mu\psi+Q^{\alpha\beta}h_{\alpha\beta}\psi\right]\nn
&=S_1^{\psi^2}+S_1^{h\psi},
\end{align}
where we integrated by parts the last term in the first integral and defined the useful tensors
\begin{equation}
 P^{\alpha\beta\mu}=\frac{\alpha\kappa^2-\omega}{\alpha\kappa}\left[\frac{\delta^{\alpha\beta}\partial^\mu\phi}{2}-\delta^{\mu(\alpha}\partial^{\beta)}\phi\right]\,,\qquad
 Q^{\alpha\beta}=\left[\left(\frac{\gamma\kappa}{4}-\frac{\omega}{2\alpha\kappa}\right)\Box\phi~\delta^{\alpha\beta}
 +\frac{\omega}{\alpha\kappa}\partial^\alpha\partial^\beta\phi\right]\,.
\end{equation}
As previously stated, due to Wick's theorem, mixed terms involving odd graviton-scalar interactions vanish, $\langle \psi^2\cdot h\psi\rangle=0$, so 
that the final expression to be evaluated will be 
\begin{align}
 \langle S_1(x)S_1(x')\rangle=\langle S_1^{\psi^2}(x)S_1^{\psi^2}(x')\rangle+\langle S_1^{h\psi}(x)S_1^{h\psi}(x')\rangle\,.
\end{align}
Let us calculate separately the two contributions in the above expression.
Taking into account all the possible contractions of the fields and hence eventual extra combinatorics factors, employing dimensional regularisation and taking the limit of $n\rightarrow4$, we have\footnote{Hereafter a prime will denote differentiation with respect to $x'$.}
\begin{align}\label{gravscal2}
\langle S_1^{\psi^2}(x)S_1^{\psi^2}(x')\rangle&=\nonumber \\
&\hspace{-3cm}=\frac{1}{M^6}\!\dint d^nx d^nx'\! \left[\frac{1}{2}\Box\phi(x)\Box'\phi(x')[\partial_\mu\partial'_\nu G(x,x')]^2
+\partial^\mu\phi(x)\partial'^\nu\phi(x')\partial_\mu\partial'_\nu G(x,x') \Box\Box' G(x,x')+\right.\nn
&\hspace{-3cm}\left.+\partial^\mu\phi(x)\partial'^\nu\phi(x')\partial_\mu \Box'G(x,x') \partial'_\nu\Box G(x,x')
+2 \partial^\mu\phi(x)\Box'\phi(x')\partial^\mu \partial'^\nu G(x,x') \partial'_\nu\Box G(x,x')\right]=\nonumber\\
&=\frac{a_L}{M^6}\int d^nx\cdot \left(\frac{1}{8}\phi\Box^{(4)}\phi-\frac{5}{4}m^2_\Lambda\phi\Box^{(3)}\phi
+\frac{15}{4}m^4_\Lambda\phi\Box^{(2)}\phi\right),
\end{align}
\begin{align}\label{gravscal}
 \langle S_1^{h\psi}(x)S_1^{h\psi}(x')\rangle&=\nonumber\\
 &\hspace{-3.2cm}=\!\!\dint \!  d^nx d^nx' \! \left[ P^{\alpha\beta\mu}(x)P^{\rho\sigma\nu}(x')G_{\alpha\beta\rho\sigma}(x,x')\partial_\mu\partial'_\nu G(x,x')\!
 +\!Q^{\alpha\beta}(x)Q^{\rho\sigma}(x') G_{\alpha\beta\rho\sigma}(x,x') G(x,x')\right.\nonumber\\
 &\left. +2 P^{\alpha\beta\mu}(x)Q^{\rho\sigma}(x') G_{\alpha\beta\rho\sigma}(x,x') \partial_\mu G(x,x')\right]=\nonumber\\
 &\hspace{-2cm}=a_L\int d^nx \left[\left(\frac{\gamma\omega}{4}-\frac{3\gamma^2}{8}+\frac{\alpha\kappa^2\gamma^2}{8}+\frac{3\gamma}{4}-\frac{\alpha\kappa^2\gamma}{2}+
 \frac{\omega}{2}\right)\kappa^2\phi\Box^{(2)}\phi\right.\nonumber\\
 &\hspace{1cm}\left.+\left(\omega^2\lambda+m^2_\Lambda\omega-\frac{m^2_\Lambda\omega^2}{2\alpha\kappa^2}-2\alpha\kappa^2\lambda\omega-\frac{\alpha\kappa^2 m^2_\Lambda}{2}+
 \alpha^2\kappa^4\lambda\right)\kappa^2\partial_\mu\phi\partial^\mu\phi\right],
\end{align}
where the operator $\Box^{(k)}\phi$ is defined as
\be
\Box^{(k)}\phi \equiv \underbrace{\Box\Box\cdots\Box}_{\textrm{$k$ times}}\phi.
\ee 
Note that all the three pieces of the first integral in \eqref{gravscal} carry terms proportional to $\alpha^{-1}$; interestingly enough most of them 
cancel each other, apart for the $-\frac{m^2\omega^2}{2\alpha}$ in the last line; this seems to be rather counter-intuitive as we expect the regularity
and finiteness of the resulting effective action when the DeWitt--gauge limit is taken, $\alpha\rightarrow0$. However, as we will show in the next
section, this term will exactly cancel the other term proportional to $\alpha^{-1}$ coming from the $\langle S_2(x,x)\rangle$ piece, guaranteeing the absence
of singular terms as $\alpha \to 0$.

All these results can be easily translated in the language of Feynman diagrams, as illustrated in Fig.~\ref{Granma}. The first two diagrams of the 
first line refer to the scalar and graviton loops of the tadpole-type. It is important to note that, as we already showed in \eqref{graviton}, in four 
dimensions the graviton tadpole--type term does not contribute to the logarithmically divergent part; however, it contributes to the finite part of the 
1-loop effective action in arbitrary dimensions, so we include it as a possible contributor. The remaining diagrams of the first line are the 1-loop 
scalar--graviton interaction terms, with full dots indicating the number of derivatives acting on background fields and internal propagators. Finally, the last line
contains the diagrams associated with the contribution from the scalar--scalar interaction.

\subsection{1--loop effective action and the structure of counter-terms}

With the results of the previous subsections in hand, we are now in the position to evaluate the 1--loop effective action \eqref{Pairings0}; using \eqref{scalar}, \eqref{gravscal2} and \eqref{gravscal} in the limit of $n\rightarrow4$ we find,
\begin{align}
\Gamma^{\text{1--loop}}&=\Braket{S_2(x,x)}- \frac{1}{2}\Braket{S_{1}(x) S_{1}(x')}\nonumber\\
&\stackrel{\tiny{\alpha\rightarrow0}}{=} a_L \cdot \int d^4 x \Bigg\{-\frac{1}{16M^6}\phi\Box^{(4)}\phi    \; +\frac{5m_\Lambda^2}{8M^6}\phi\Box^{(3)}\phi + \; \nonumber\\
&\hspace{1cm}+ \phi\Box^{(2)}\phi \left[ \frac{\kappa^2}{4} \cdot \left( \alpha\kappa^2\gamma+\frac{3\gamma^2}{4}-\frac{3\gamma}{2}-\frac{\alpha\kappa^2\gamma^2}{4}-\omega-\frac{\gamma\omega}{2} \right) -\frac{15m_\Lambda^4}{8M^6}\right]+\nonumber\\
&\hspace{2cm} + \partial_\mu\phi\partial^\mu\phi \cdot \frac{\kappa^2}{2} \cdot \left[\frac{\gamma m_\Lambda^2}{8} -  \lambda\omega^2  -  \omega m_\Lambda^2 +2\alpha\kappa^2\lambda \omega+\frac{\alpha \kappa^2 m_\Lambda^2}{2}- \alpha^2\kappa^4\lambda\right]\Bigg\}. \label{Eff-action}
\end{align}

Note that although the limit $\alpha\rightarrow0$ has to be taken to recover the gauge-independent result, we left in the final expression 
all the $\alpha$-dependent terms for ease of comparison with gauge--dependent calculations. We further stress that, in four dimensions one has $\lambda\stackrel{\tiny{n\rightarrow4}}{=}\Lambda$. It is also worth noting that 
if the cosmological constant vanishes, then also the effective mass of purely geometrical origin $m_\Lambda$ will do so. 
The book--keeping parameter $\gamma$ that highlights the connection--dependent part, should be set to unity for the covariant result to be recovered.
Notice also that, even in the 
limiting case $\gamma=0$, that is when the contribution coming from the field--space connection is switched off and one is back to the usual implementation
of the background--field method, parts of the coefficients in \eqref{Eff-action} still depend on the sole gauge parameters $\alpha$ and $\omega$. This should be interpreted as a possible warning sign for any
calculation that, fixing {\em ab initio} anyone among the parameters $\gamma$, $\omega$ or $\alpha$, can in principle arrive to gauge dependent results.

The logarithmic divergence, tracked by the $a_L$ factor in \eqref{Eff-action}, suggests a straightforward expression of the counter-terms to be added to the bare action in order to renormalise it. An important point that should be noted is the purely quantum-gravitational nature of the contribution to the renormalisation of the kinetic term (last line of \eqref{Eff-action}), which comes exclusively from the graviton-scalar interaction term and from the scalar tadpole term which' origin in turn resides in the field--space connection. This does not come as a surprise and is also a useful counter-check of the consistency of the calculation, since it is well--known that there is no wave--function renormalisation at 1--loop from scalar loops. 

On the other hand, the three terms contained in the first two lines of the effective action result from Galileons and gravitons running in a loop, in particular, they come from both the scalar--scalar and the graviton--scalar interactions, as can be seen from \eqref{gravscal2} and \eqref{gravscal}. We emphasise here the novelty of the two operators, $\sim \frac{m_\Lambda^2}{M^6}\phi\Box^{(3)}\phi$ and $\sim -\frac{m_\Lambda^4}{M^6}\phi\Box^{(2)}\phi$. {\it They represent genuinely new operators, which' origin is directly related to the gravitational interactions and the geometrically--induced mass $m_\Lambda$}. It is also intriguing that their relevance is controlled by a ratio of the cosmological vacuum and the Galileon mass scale respectively \footnote{Note that in the standard (non-covariant) approach within the cubic Galileon theory, the only visible operator through dimensional regularisation is $\phi\Box^{(4)}\phi$ \cite{dePaulaNetto:2012hm}. The terms $\sim \phi\Box^{(2)}\phi$ and $\sim \phi\Box^{(3)}\phi$ have been found previously in this context using cut-off regularisation in the standard approach \cite{Brouzakis:2013lla}, representing quartic and quadratic divergences respectively, and with their amplitude controlled by an appropriate power of the cut--off scale.}.
If included in the original bare action, care is required so that they do not produce unwanted instabilities associated with the higher--order nature of these operators, in particular within scenarios where the Galileon field is responsible for driving the accelerated expansion of the Universe.

\section{Discussion and concluding remarks} \label{sec:Conclusions}
In this work, we presented for the first time an explicit exposition of quantum corrections for a Galileon theory, that accommodates for both scalar and gravitational quantum--mechanical effects, revealing new interesting features. Following our previous work \cite{Saltas:2016nkg}, we employed the powerful field--re-parametrisation invariant formalism of the covariant effective action at 1--loop. The consideration of a gauge theory, makes the requirement of gauge--invariant results a highly desirable feature and our final results are free from such ambiguities.  
\\

{\it The key points of our analysis can be summarised as follows:}
\\

1. Although quantum--gravitational effects are expected to occur at energies close or above the Planck scale, our results suggest that the presence of non--trivial derivative interactions in the bare action, and within a covariant framework, lead to new gravitationally--induced interactions at 1--loop, potentially relevant at scales as large as the cosmological horizon. This is intimately related to the fact that covariance in field--space requires that scalar and gravitational fluctuations are treated on equal footing, leading to novel features at the perturbative level, unseen within non--covariant calculations. (See also the discussion around equations (\ref{Explicit-connection}) and (\ref{eq:Scalar-prop}) in the main text.) 
\\

2. The main result of our computation is the structure of 1--loop divergences, as presented in the expression (\ref{Eff-action}). From there, the extraction of the relevant counter-terms and the renormalised effective action is a straightforward task.  What is more, from the effective action (\ref{Eff-action}) it can be seen that the wave--function renormalisation of the theory receives purely quantum--gravitational contributions, a result previously known within other scalar--field theory contexts at 1--loop. 
\\

3. Most importantly, the 1--loop effective action (\ref{Eff-action}) suggests the new terms required to be added in the original bare action, for the UV--divergences to be consistently absorbed. The new operators revealed by our analysis correspond to higher--derivative interactions for the Galileon described by, 
\be
\sim \frac{\Lambda}{M^6}\phi\Box^{(3)}\phi, \; \; \sim -\frac{\Lambda^2}{M^6}\phi\Box^{(2)}\phi.
\ee
They are both a direct result of the presence of the mass-type interaction, with $m^2_{\Lambda} \sim \Lambda$, which the Galileon field developed at the perturbative level, and is unseen within non--covariant frameworks. They represent genuine logarithmic divergences of gravitational origin, and their relevance is controlled by the ratio of the cosmological vacuum $\Lambda$ to the Galileon scale $M$ to an appropriate power. It this sense, they can be understood as quantum--gravity induced corrections, which could potentially manifest themselves even at infrared scales. In this context, care has to be taken for them to remain subdominant in cosmological applications, an analysis which we leave for a future study.\\

Working at quadratic order in the background scalar field $\phi$, our results did not capture the (non--) renormalisation of the cubic Galileon term. However, its presence already manifested itself through its impact on lower--order Galileon operators as we discussed earlier. It would be an interesting, yet very demanding task, to perform a similar analysis at cubic order in the background field, including the back-reaction of gravity, aiming to reveal a richer and possibly more surprising features within this context. What is more, as discussed earlier, the cosmological relevance of our results, in particular the significance of 1--loop corrections for the scenario where the Galileon field inflates the early-- or the late--time Universe, is yet another interesting research direction which we leave for the future. 

\acknowledgments
VV thanks the Funda\c c\~ao para a
Ci\^encia e Tecnologia (FCT)-Portugal for the financial support provided  through the grants SFRH/BPD/77678/2011 and  
UID/FIS/00099/2013. IDS is supported by FCT under the grant SFRH/BPD/95204/2013, and further acknowledges UID/FIS/04434/2013 and the project FCT-DAAD 6818/2016-17. The authors thank Nino Flachi, Tony Padilla, Iggy Sawicki and Dimitri Skliros for useful discussions. They are also grateful to Tim Morris for constructive criticism and feedback.

\appendix
\section{Some more explicit calculations}
In this section we present some helpful intermediate steps in the expansion of the action up to second order in field fluctuations. 
For the expansion of the curvature sector of the action $S$ we proceed as follows,
\begin{align} 
 \delta^{(2)}_{g g} S_{\text G}  & =   - \frac{2}{\kappa^2}  \int d^4 x  \sqrt{g} \left( 2 \frac{ \delta \sqrt{g}}{ \sqrt{g}} \cdot  \delta R  + \delta^{(2)}R \right)   \nonumber \\
& = -  2 \int d^4 x  \sqrt{g}  \Bigg\{  h \partial_\alpha \partial_\beta h^{\alpha \beta} - \frac{1}{2}h \Box h + \frac{1}{2} h^{\alpha \beta} \Box h_{\alpha \beta} + \partial_\beta h^{\alpha \beta} \partial_\lambda h^{\lambda}{}_{\alpha}
\Bigg \},
\end{align}
using $\bar R \equiv R(\bar{g}_{\mu \nu} = \delta_{\mu \nu}) = 0$.
In a similar fashion, the expansion of the matter part of the action organises in matrix form as
\be
\delta_{ij}^{(2)}S_{\text{M}} = \left( \begin{array}{cc}
\delta_{gg}^{(2)}S_{\text{M}} \; \; & \delta_{g\phi}^{(2)}S_{\text{M}} \\
& \\
\delta_{g \phi}^{(2)}S_{\text{M}} \; \; & \delta_{\phi \phi}^{(2)}S_{\text{M}}
 \end{array} \right),
\ee
with the respective entries given by, 

\begin{align}
 \delta^{(2)}_{\phi \phi} S_{\text M}  & =   \int d^4 x  \sqrt{g} \left\{ 
 (1+ B)  \cdot \delta^{(2)}_{\phi \phi} X     + 2 \cdot \delta_\phi X \cdot \delta_\phi B
\right\} \nonumber \\
& =  \int d^4 x  \sqrt{g} \left\{ 
(1+B) \cdot  \psi_\mu \cdot    \psi^\mu   + 2M^{-3}  \cdot  \phi_\mu \cdot  \psi^\mu\cdot \Box \psi
\right\},
\end{align}


\begin{align}
 \delta^{(2)}_{g \phi} S_{\text M} = \kappa \int d^4 x \sqrt{g}  &\cdot \left\{ 
    \frac{1}{2}(1+B)\cdot  \phi_\alpha h \psi^{\alpha} + \frac{1}{2}M^{-3} X\cdot  h \Box \psi 
- (1+B)\phi_\alpha \cdot h^{\alpha \beta} \psi_{\beta}  \right.    \nonumber \\
& \hspace{-1cm} - \frac{1}{2}M^{-3} \phi_\alpha \phi_\beta \cdot h^{\alpha \beta} \Box \psi  - M^{-3} \phi^\mu \phi_{\alpha \beta} h^{\alpha \beta}  \psi_{\mu}  - M^{-3} \phi^\mu \phi^\kappa \partial_\beta h^{\beta}_{\kappa} \psi_{\mu}  \nonumber \\
& \hspace{-1cm} \left.+ \frac{1}{2}M^{-3} \phi^\mu \phi^\kappa \partial_{\kappa}h \psi_{\mu} + M^{-3}X \left( - h^{\alpha \beta}  \psi_{\alpha \beta} -  \partial_\beta h^{\beta}_{\kappa} \psi^\kappa  +  \frac{1}{2}\partial_\kappa h \psi^\kappa \right)
\right\},
\end{align}

\begin{align}
& \delta^{(2)}_{g g} S_{ \text M} = \kappa^2 \int d^4 x  \sqrt{g} \left\{  \frac{1}{2} X(1+B) \cdot \left( \frac{1}{2} h^2 - h_{\mu \nu}h^{\mu \nu} \right)  + (1+B)\,\phi_\alpha \phi_\beta \left(-\frac{1}{2}h^{\alpha \beta} h + h^{\alpha \kappa} h_{\kappa}^{\beta} \right)  \right.  \nonumber \\
& \hspace{2cm} + X \phi_{\alpha \beta} \cdot \left( 2 h^{\alpha \kappa} h_{\kappa}^{\beta} - h h^{\alpha \beta} \right) - \frac{1}{2} \phi^{\kappa}(X \delta_{\mu \nu} - \phi_\mu \phi_\nu)\cdot \left( 2 h^{\mu \nu} \partial_{\beta} h^{\beta}_{\kappa}  - h^{\mu \nu}\partial_{\kappa}h \right)\nonumber \\
& \hspace{1cm}  \left. + \phi_{\mu} \phi_{\nu} \phi_{\alpha \beta} h^{\mu \nu} h^{\alpha \beta} 
+ \frac{1}{2} X \phi^{\kappa} \left( 2h^{\alpha \beta}\partial_{\alpha}h_{\beta \kappa} - h^{\alpha \beta} \partial_{\kappa}h_{\alpha \beta} \right) 
+ X \phi_\rho \left(2h^{\rho \kappa} \partial^{\alpha}h_{\alpha \kappa}  - h^{\rho \kappa} \partial_{\kappa}h \right)
\right\}\nonumber\\
%
&= \kappa^2 \int d^4 x  \sqrt{g}  \left\{  \frac{1}{2} X \left( \frac{1}{2} h^2 - h_{\mu \nu}h^{\mu \nu} \right)     
+ \phi_\alpha \phi_\beta \left(-\frac{1}{2}h^{\alpha \beta} h + h^{\alpha \kappa} h_{\kappa}^{\beta} \right) \!\right\} \nonumber \\
%
& + \kappa^2 \int d^4 x \sqrt{g} \left\{  \frac{1}{2} X B  \left( \frac{1}{2} h^2 - h_{\mu \nu}h^{\mu \nu} \right)  
 + B \phi_\alpha \phi_\beta  \left(-\frac{1}{2}h^{\alpha \beta} h + h^{\alpha \kappa} h_{\kappa}^{\beta} \right)  + X \phi_{\alpha \beta}  \left( 2 h^{\alpha \kappa} h_{\kappa}^{\beta}\right.\right.\nonumber\\
 &\hspace{2cm} \left.- h h^{\alpha \beta} \right)  - \frac{1}{2} \phi^{\kappa}(X \delta_{\mu \nu} - \phi_\mu \phi_\nu) \left( 2h^{\mu \nu} \partial_{\beta} h^{\beta}_{\kappa}  - h^{\mu \nu}\partial_{\kappa}h \right)  + \phi_{\mu} \phi_{\nu} \phi_{\alpha \beta}  h^{\mu \nu} h^{\alpha \beta} \nonumber \\
& 
\hspace{2cm}\left. + \frac{1}{2} X \phi^{\kappa} \cdot \left( 2h^{\alpha \beta}\partial_{\alpha}h_{\beta \kappa} - h^{\alpha \beta} \partial_{\kappa}h_{\alpha \beta} \right) 
+ X \phi_\rho \cdot \left(2h^{\rho \kappa} \partial^{\alpha}h_{\alpha \kappa}  - h^{\rho \kappa} \partial_{\kappa}h \right)
\right\}.
\end{align}

\bibliographystyle{utcaps}
\bibliography{VG.bib}

\end{document}